\documentclass{ifacconf}
\pdfoutput=1

\usepackage{xfrac} 
\usepackage{booktabs}
\usepackage{siunitx}
\usepackage{pgf}  %
\usepackage{pgfplots}
\usepackage{multirow}
\usepackage{array}
\usepackage{caption}
\usepackage{adjustbox}
\usepackage{amsmath, nccmath}
\usepackage{amssymb}
\usepackage{graphicx}      
\usepackage{natbib}        
\usepackage[hyphens]{url}
\usepackage{threeparttable}

\begin{document}
\begin{frontmatter}

\title{Data Center Chiller Plant Optimization via Mixed-Integer Nonlinear Differentiable Predictive Control}

\thanks[footnoteinfo]{This research was supported by the U.S. Department of Energy Building Technologies Office (BTO), USA within the Office of Energy Efficiency and Renewable Energy (EERE), USA under Contract No. DE-AC05-76RL01830. J\'an Drgo\v na was supported by the Ralph O’Connor Sustainable Energy Institute at Johns Hopkins University. J.B. and M.G. also acknowledge the contribution of the Slovak Research and Development Agency under the project APVV-22-0436, and the Scientific Grant Agency of the Slovak Republic under the grant 1/0401/26.}

\author[First,Second]{J\'an Boldock\'y} 
\author[Second]{Cary Faulkner}
\author[Second]{Elad Michael} 
\author[First]{Martin Gulan}
\author[Second]{Aaron Tuor}
\author[Third,Second]{J\'an Drgo\v na}

\address[First]{Slovak University of Technology, 
   Bratislava, Slovakia \\(e-mail:
   \{jan.boldocky, martin.gulan\}@stuba.sk)}
\address[Second]{Pacific Northwest National Laboratory, Richland, WA 99352 USA (e-mail: \{cary.faulkner, elad.michael, aaron.tuor\}@pnnl.gov)}
\address[Third]{Johns Hopkins University, MD 21218, USA, \\(e-mail: jdrgona1@jh.edu)}

\begin{abstract}                
We present a computationally tractable framework for real-time predictive control of multi-chiller plants that involve both discrete and continuous control decisions coupled through nonlinear dynamics, resulting in a mixed-integer optimal control problem. To address this challenge, we extend Differentiable Predictive Control (DPC)---a self-supervised, model-based learning methodology for approximately solving parametric optimal control problems---to accommodate mixed-integer control policies.  We benchmark the proposed framework against a state-of-the-art Model Predictive Control (MPC) solver and a fast heuristic Rule-Based Controller (RBC). Simulation results demonstrate that our approach achieves significant energy savings over the RBC while maintaining orders-of-magnitude faster computation times than MPC, offering a scalable and practical alternative to conventional combinatorial mixed-integer control formulations.
\end{abstract}


\end{frontmatter}

\section{Introduction}

The rapid expansion of artificial intelligence and digital infrastructure has led to a sharp increase in global data center electricity demand. 
In the United States alone, data centers currently consume about $150$~TWh of electricity annually and are projected to grow to between $200$~TWh and $500$~TWh by 2030~\citep{aljbour2024powering}. 
With this expansion comes an accompanying demand for high-efficiency cooling systems in data centers.
Chiller plants are commonly deployed to perform this task. A chiller plant typically includes heat pumps (often referred to as chillers), pumps, and cooling towers, with the chillers being the most energy-intensive components within the plant.
The controls of chiller plants involve both continuous decision variables (e.g., chilled water supply temperature, differential pressure) and discrete decision variables (e.g., number of chillers active, pump mode).
Optimizing these coupled decisions can significantly reduce electricity consumption, ensure reliable operation, and extend the lifespan of costly mechanical assets.
However, this requires the solution to a high-dimensional mixed-integer nonlinear optimal control problem, which is computationally challenging to solve in real time.

The most widely used industrial strategy is Rule-Based Control (RBC), valued for its interpretability and low implementation cost.
Several works have focused on refining RBC through heuristic tuning and performance optimization.
For example, \citet{ma2011supervisory} developed a supervisory RBC framework that stages chillers based on the part-load ratio (PLR) to improve efficiency;
\citet{huang2016amelioration} proposed adaptive PLR thresholds to balance energy use and load following;
\citet{liu2017optimal} formulated rule-based optimization of chiller sequencing using plant-level performance metrics;
and \citet{faulkner2025development} recently performed simulations optimizing control parameters of RBC strategies to balance chiller, pump, and tower loads.
While RBC methods offer fast and interpretable operation, they rely on fixed heuristics and typically fail to account for future disturbances, often resulting in suboptimal energy efficiency or constraint violations.

To overcome these limitations, researchers have investigated predictive control strategies for chiller plants~\citep{sala2020predictive, terzi2020learning, chan2022development, fan2023model, he2023predictive}, that leverage dynamic models and load forecasts to proactively schedule chillers and optimize continuous setpoints.
For example, Model Predictive Control (MPC) has been shown to improve chiller staging under varying ambient and load conditions~\citep{sala2020predictive, he2023predictive}, while data-driven or learning-based predictive control variants adapt to plant-specific nonlinearities~\citep{terzi2020learning, chan2022development, fan2023model}. \textcolor{black}{
However, most existing methods handle either continuous~\citep{pan2024nonlinear} or discrete control actions in isolation, and few address the mixed-integer nature of chiller plant operation.}

A principled way to unify discrete and continuous decisions under constraints is Mixed-Integer Model Predictive Control (MI-MPC)~\citep{mcallister2022advances, dua2002, Richards2005, Kirches2011, Takapoui02012020}.
MI-MPC provides a systematic framework for incorporating nonlinear dynamics and hard constraints while producing high-quality solutions.
However, its computational complexity grows combinatorially with the horizon length and the number of integer variables, making real-time operation intractable, particularly for large-scale systems such as multi-chiller plants.
Approximation strategies such as move blocking, surrogate linearizations, or warm-starts help alleviate but do not eliminate these scaling limitations, especially when tight sampling periods and long prediction horizons are required. \textcolor{black}{A representative example in the context of central chiller plants is the decomposition-based MPC framework of \cite{deng2015}, which combines dynamic programming and mixed-integer linear programming to obtain tractable suboptimal solutions; however, the approach still requires repeated online optimization, limiting the computational scalability.}

These challenges have motivated a growing body of research on learning-based alternatives that avoid online combinatorial search. In imitation-based or approximate MPC approaches, machine learning (ML) policies are trained to mimic optimal or near-optimal mixed-integer solutions offline and then deployed as explicit controllers, offering fast real-time evaluation while maintaining reasonable constraint satisfaction\citep{karg2018,cauligi2020,DOMAHIDI2014763}. 
In parallel, warm-start techniques utilize ML on previously solved MIP instances to initialize subsequent optimizations with good primal and dual guesses, reducing branch-and-bound exploration and improving real-time feasibility in receding-horizon settings \citep{masti2019,marcucci2021,reiter2024}.
 Beyond imitation learning, the broader field of learning to optimize (L2O) has also explored data-driven acceleration of MIP solvers through learned branching, node selection, cutting, and heuristic policies that reduce solve times by orders of magnitude~\citep{Bertsimas2022,tang2025, he2014learning, zarpellon2021parameterizing}. These developments motivate a self-supervised Differentiable Predictive Control framework, which merges the interpretability of MPC with the scalability of L2O.

To address these challenges, we build on our prior work on Differentiable Predictive Control (DPC)~\citep{drgona2022differentiable,Drgona_DPC2024}, which formulates parametric MPC problems as differentiable programs, enabling offline self-supervised learning of explicit neural control policies through gradient-based learning.
Building upon this foundation, our recent work on Mixed-Integer DPC (MI-DPC)~\citep{boldocky2025learning} extended this idea to handle mixed-integer optimal control problems (MI-OCP) with discrete decision variables. 
In this work, we demonstrate the application of MI-DPC to a challenging multi-chiller plant optimization problem characterized by nonlinear dynamics, mixed-integer decisions, subject to input and state constraints.
The proposed MI-DPC approach provides a scalable self-supervised learning algorithm for synthesizing real-time capable explicit neural control policies that yield high-quality approximate solutions to the underlying MI-OCP,  without online combinatorial search.

The contributions of this work include: 
(i) a nonlinear control-oriented multiple chiller system model;
(ii) extension of the MI-DPC framework to mixed-integer nonlinear optimal control problems arising in chiller plant operations;
(iii) extension of the MI-DPC framework to handle binary decision variables and the introduction of the binary-variance regularization technique, necessary to prevent high-frequency switching behavior;
(iv) a systematic performance comparison of the proposed MI-DPC against an RBC and an MPC solved with a state-of-the-art mixed-integer solver;
and (v) the release of open-source code
\footnote{Code is available at github.com/pnnl/MI-DPC.}
enabling replication of the numerical experiments.

\section{Problem Formulation}\label{sec:prob_form}
\begin{figure}[h!]
\centering
\includegraphics[width=0.99\columnwidth]{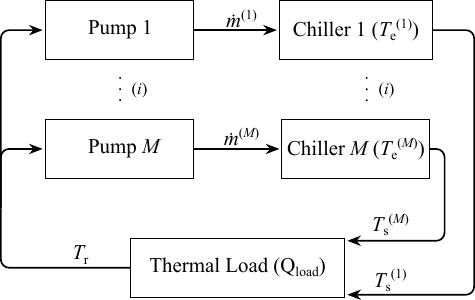}
\caption{Schematic of a multiple-chiller plant arranged in a parallel configuration.\label{fig:chiller-diagram}}
\end{figure}

We consider a simplified multi-chiller plant, illustrated in Figure~\ref{fig:chiller-diagram}, with chillers operating in parallel.
In the following, we use a superscript to denote an association of a variable with chiller loop $i$, i.e. $\dot{m}^{(1)}$ represents the mass flow through the first chiller. We further denote quantities sharing the same units by a common symbol with a subscript, such as $T_\mathrm{e},T_\mathrm{s},T_\mathrm{r}$ for the evaporator, supply and return temperatures, respectively. In the following, we use $M$ to denote the number of chillers in the system.

\subsection{System Model}\label{sec:system_description}

The continuous-time thermal dynamics of the multi-chiller system are
\begin{subequations}
\begin{fleqn}
\begin{align}
Q^{(i)}(t) &= \eta_\mathrm{r}\dot{m}^{(i)}(t)\; c_\mathrm{p}\delta^{(i)}(t) \left( T_\mathrm{r}(t)-T^{(i)}_\mathrm{s}(t) \right), \label{eq:cooling_delivered}\\ 
\dot{T}_\mathrm{r}(t) &= \frac{1}{C_\mathrm{r}}\left(\tilde{Q}_\text{load}(t) - \sum_{i=1}^{M}Q^{(i)}(t) \right), \label{eq:temp_return}\\
\dot{T}_\mathrm{s}^{(i)}(t) &= -\frac{c_\mathrm{p} \eta_\mathrm{s}}{C^{(i)}}\left(\delta^{(i)}(t)\dot{m}^{(i)}(t) \left(T^{(i)}_\mathrm{s}(t)-T^{(i)}_\mathrm{e}(t)\right)\right),\!\! \label{eq:temp_supply}
\end{align}
\end{fleqn}
\end{subequations}
where $Q^{(i)}(t)\!\in\!\mathbb{R}_{\ge0}$ is the delivered cooling from chiller $i$ at time $t$, $\delta^{(i)}(t)\!\in\!\{0,1\}$ denotes chiller's on/off state, $c_\mathrm{p}\! \in\!\mathbb{R}_{>0}$ is the specific heat capacity of the coolant, and $C_\mathrm{r},C^{(i)}\!\in\!\mathbb{R}_{>0}$ are the thermal capacitances of the return and chiller loops. Moreover, $\eta_\mathrm{r},\eta_\mathrm{s}\!\in\!(0,1)$ denote the heat-exchanger efficiency coefficients, and $\tilde{Q}_\text{load}(t)\!\in\! \mathbb{R}_{\ge0}$ is the effective thermal load, estimated using a low-pass, discrete-time finite impulse response filter
\begin{equation}
\tilde{Q}_\text{load} (t) = \sum_{l=0}^{L} h_l Q_\text{load}(t-l\Delta t),
\end{equation}
where $Q_\text{load}\in\mathbb{R}_{\ge 0}$ is the direct server load---an exogenous variable in this context, $\Delta t$ denotes the length of a discrete-time step, $L\!\in\!{N}_{\ge1}$ is the filter order, and $(h_l)_{l=0}^{L}\!\in\!\mathbb{R}_{>0}$ are the filter coefficients, with $\sum_l h_l\!=\!1$. \textcolor{black}{Note that $Q_\mathrm{load}$ is treated as a deterministic known variable in our study. We consider the FIR filter to capture higher-order heat-transfer dynamics, thereby enhancing the model's expressive capacity. Naturally, the filter coefficients can be a subject of system identification along with other system parameters.}

The power consumption of chillers $P^{(i)}_\text{chiller}(t)\in\mathbb{R}_{\ge 0}$ and pumps $P^{(i)}_\text{pump}(t)\in\mathbb{R}_{\ge 0}$ are modeled as
\begin{subequations}\label{eq:power-equations}
\begin{align}
P^{(i)}_\text{chiller}(t) &= \frac{Q^{(i)}(t)}{\text{COP}^{(i)}(t)}+\rho^{(i)}\delta^{(i)}(t) \label{eq:chiller_power},\\
\text{COP}^{(i)}(t) &= a_0^{(i)}+a_1^{(i)}\frac{Q^{(i)}(t)}{Q_\text{max}^{(i)}}+a_2^{(i)}\left(\frac{Q^{(i)}(t)}{Q_\text{max}^{(i)}}\right)^2,\\
P^{(i)}_\text{pump}(t) &= \gamma^{(i)}\left( \dot{m}^{(i)}(t)\delta^{(i)}(t) \right)^3, \label{eq:pump_power}
\end{align}
\end{subequations}
for the system constants $a_{0:2}^{(i)},\gamma^{(i)},\rho^{(i)},Q_\text{max}^{(i)}\!\in\!\mathbb{R}$. The constant $\rho$ represents the base power required for chiller operation. In this formulation, the chiller's coefficient of performance $\text{COP}^{(i)}(t)\in\mathbb{R}_{\ge0}$ is a function of the part load ratio (PLR), the ratio of current cooling to maximum cooling capacity. Ensuring that each chiller is operating near its highest efficiency load ratio is the core of the underlying optimal control problem.
In the following, we discretize the continuous time dynamics given by \eqref{eq:temp_return} and \eqref{eq:temp_supply} using a fourth-order Runge-Kutta method with a time-step length $\Delta t$ and discrete time index $k$.

\subsection{Optimal Control Problem} \label{ssec:opt_control}
Let $u^{(i)}_k\!:=\!\{\delta^{(i)}(t),T_\mathrm{e}^{(i)}(t),\dot{m}^{(i)}(t)\}$ denote the set of control variables and $x^{(i)}_k\!:=\!\{T_\mathrm{s}^{(i)}(t),T_\mathrm{r}(t)\}$ denote the set of state variables at discrete time step $k$.
Then the optimal control problem (OCP) can be formulated as follows:
\begin{subequations} \label{eq:ocp}   
\begin{align}\nonumber
\min_{\{u^{(i)}_{k}\}_{i=1}^{M}{,}_{k=0}^{N\!-\!1}} \sum^{N-1}_{k=0}\sum^M_{i=1}  & P^{(i)}_{\text{chiller},k}+P^{(i)}_{\text{pump},k}+  \\
 \|\Delta\delta^{(i)}_{k}\|_R^2 + & \lambda_W\left(Q_{k}\!-\!Q_{\text{load},k}\right)^2 \label{eq:ocp-objective}\\
\noalign{\text{s.t.}} 
x^{(i)}_{k+1}=&f\left(x^{(i)}_{k}, u^{(i)}_{k},\tilde{Q}_{\text{load},k}, \Delta t\right), \label{eq:ocp-dynamics}\\
Q_{k}  = \sum_i Q^{(i)}_{k},\; &\forall k\in\{0,\dots,N\!-\!1\},\\
 \sum_i\delta^{(i)}_{k}\ge \;1,\;\, &\forall k \in \{0,\dots,N\!-\!1\}, \label{eq:ocp-oneon}\\
u^{(i)}_{\min}\leq u^{(i)}_{k}\leq \;u^{(i)}_{\max},\; &\forall k \in \{0,\dots,N\!-\!1\},\label{eq:ocp-input-constraints}\\
x^{(i)}_{\min}\leq x^{(i)}_{k}\leq \;x^{(i)}_{\max},\;\,\, &\forall k \in \{0,\dots,N\}, \label{eq:ocp-state-constraints}\\
Q^{(i)}_{k} \leq \;Q^{(i)}_{\max},\;\, &\forall k \in \{0,\dots,N\!-\!1\}, \label{eq:cooling-constraint}\\
\eqref{eq:chiller_power}-\eqref{eq:pump_power},\;\,\, &\forall k \in \{0,\dots,N\!-\!1\}\label{eq:ocp-power-constraints},\\
\delta^{(i)}_{k} \in \;\{0,1\}, \;\, &\forall k \in \{0,\dots,N\!-\!1\},\label{eq:ocp-binary}\\
x^{(i)}(0) = &\;x^{(i)}(t).
\end{align}
\end{subequations}
Here, the objective function \eqref{eq:ocp-objective} penalizes the chiller and pump power consumption, switching of chiller on/off status $\Delta\delta^{(i)}_{k}\!:=\!\delta^{(i)}_{k+1}\!-\!\delta^{(i)}_{k} \;\forall k\!\in\!\{0,\dots, N\!-\!2\}$---weighted by a diagonal matrix $R$ of size $M$, and the squared error between total delivered cooling $Q\!\in\!\mathbb{R}_{\ge0}$ and the thermal load demand, weighted by coefficient $\lambda_W\!\in\!\mathbb{R}_{>0}$. \textcolor{black}{Weighting matrices $R$ and $\lambda_W$ directly govern the trade-off between integer switching frequency and load tracking error minimization. Larger $R$ suppresses frequent switching behavuour, while larger $\lambda_W$ values prioritizes minimizing the load tracking error.}
The dynamics \eqref{eq:ocp-dynamics} is solved using RK4 integration of the thermal equations \eqref{eq:temp_return}--\eqref{eq:temp_supply}. Constraint~\eqref{eq:ocp-oneon} ensures that at least one chiller remains active at all times, serving as a process safety constraint. Finally, constraints \eqref{eq:ocp-input-constraints}--\eqref{eq:ocp-power-constraints} enforce operational and physical systems limits. 

Note that the OCP~\eqref{eq:ocp}  is non-trivial, primarily due to system dynamics, where the binary decision variable is coupled with two bilinear terms in \eqref{eq:temp_return} and \eqref{eq:temp_supply}. Additionally, the OCP includes a set of polynomial constraints in \eqref{eq:ocp-power-constraints}, making it a mixed-integer nonlinear program.

\subsection{Rule-Based Control}
Due to the complexity of the given OCP, a rule-based control (RBC) is the most common control strategy used for improving energy efficiency of the chiller plants. Assuming the PLR--COP characteristic is known, a typical approach for multi-chiller systems is to stage the chillers based on system's PLR, defined as
\begin{equation}
    \text{PLR}_{k}\!=\! \frac{\sum_{i=1}^MQ^{(i)}_{k}}{{[Q_\text{max}^{(1)},\dots,Q_\text{max}^{(M)}][\delta^{(1)}_{k},\dots,\delta^{(M)}_{k}]^\top}
    }.
\end{equation}
In cases where the chiller plant consists of units with similar PLR--COP characteristics, a simple sequential threshold-based staging strategy can be employed:
\begin{equation}
    s_{k+1} =
\begin{cases}
\min\left(M,s_{k+1}\right), & \text{if } \text{PLR}_k\!>\!\overline{t},\\[0pt]
\max\left(1,s_{k-1}\right), & \text{if } \text{PLR}_k\!<\!\underline{t},\\[0pt]
s_{k}, & \text{otherwise,}
\end{cases}
\end{equation}
where
$
    s_{k}\!:=\!\sum_{i=1}^M\delta^{(i)}_{k}\!\in\!\mathbb{Z}_{>0}
$
denotes the number of active chillers at time $k$, and
$\underline{t},\,\overline{t}\!\in\!(0,1)$ are the staging thresholds---selected to avoid high-frequency switching while maintaining the effective operating range of the system. In addition, the continuous control variables are typically fixed at constant values that prevent process constraint violations.

\section{Methodology}
In this section, we describe the methodology designed to reformulate the defined OCP \eqref{eq:ocp} as a differentiable program suitable for the DPC framework. 
We preserve the differentiable control objective \eqref{eq:ocp-objective} and enforce constraints \eqref{eq:ocp-dynamics}--\eqref{eq:ocp-power-constraints} using either projections, where the constraints are enforced within a differentiable computational graph by projecting the variables onto their feasible sets, or relaxations using penalty terms, where the constraints are relaxed and incorporated within a loss function. The technique used to ensure the integrality of the discrete-valued decision variable is discussed in Section~\ref{sec:integrality}. 
\begin{figure*}[t]
\centering
\includegraphics[width=0.99\linewidth]{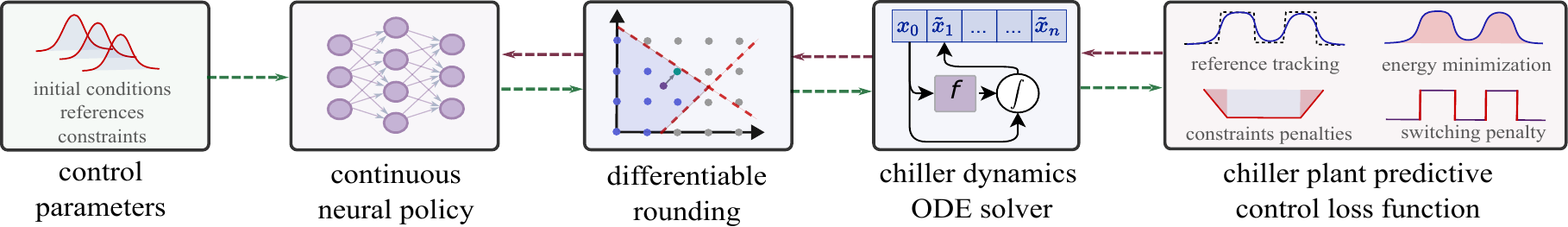}
\vspace{-0.2cm}
\caption{Conceptual diagram of the mixed-integer differentiable predictive control for nonlinear chiller plant optimization. Green dashed arrows represent the forward pass, while red dashed arrows represent the backward pass.}
    \label{fig:MI-DPC}
\end{figure*}

\subsection{Problem Reformulation}
To cast \eqref{eq:ocp} as a differentiable program, we relax the constraints \eqref{eq:ocp-input-constraints}--\eqref{eq:ocp-state-constraints} using the penalty method. Moreover, the constraint \eqref{eq:ocp-oneon} is enforced by fixing $\delta^{(2)}(k)\!=\!1$, whereas \eqref{eq:ocp-dynamics}, \eqref{eq:cooling-constraint}, and \eqref{eq:ocp-power-constraints} can be incorporated within the differentiable computational graph and thus do not require any relaxation.
Considering the above, the OCP can be recast as
\begin{subequations}
\label{eq:MI-DPC}
    \begin{align} \nonumber
\min_{\theta_1,\theta_2,\theta_3}  \!\! \!\mathbb{E}_{ \xi_k \sim P_{ \xi} }\!  
   &\bigg[\sum_{k=0}^{N-1} \sum^M_{i=1}P^{(i)}_{\text{chiller},k}\!+\!P^{(i)}_{\text{pump},k}\!+\!\|\Delta\delta^{(i)}_k\|_R^2  \\[0pt] & + \lambda_W\left(Q_k\!-\!Q_{\text{load},k}\right)^2 +\!\lambda_xq(x_k) \label{eq:dpc-objective}\nonumber\\
    & \!+\!\lambda_up(u_k) + \|\tilde\delta^{(i)}(1-\tilde\delta^{(i)})\|^2_\Lambda\bigg]\\[-10pt]
\noalign{\text{s.t.}} \nonumber\\[-15pt]
\;  x^{(i)}_{k+1}=&f\left(x^{(i)}_k, u^{(i)}_k,\tilde{Q}_{\text{load},k}, \Delta t\right), \label{eq:dpc-dynamics}\\
\dot{m}^{(i)}_k\!=&\,\pi_{\theta_1}(\xi_k),\;\\
T_\mathrm{e,k}^{(i)}\!=&\,\pi_{\theta_2}(\xi_k),\;\\ 
\tilde\delta^{(i)}_k\!=&\,\pi_{\theta_3}(\xi_k),\\
\delta^{(i)}_k \in &\;\{0,1\}, \;\, \forall k \in \{0,\dots,N\!-\!1\},\label{eq:dpc-binary}\\
\delta^{(2)}_k =& 1,\; \forall k\in\{0,\dots,N-1\}, \label{eq:dpc-oneon}\\
\xi_0\!:=\!\;[x^{(i)}(t)&, \tilde{Q}_\text{load}(t), Q_\text{load}(t),\dots,Q_\text{load}(t\!+\!N\Delta t)]^\top\!\!\!.
    \label{eq:dpc-xi}\end{align}
\end{subequations}
Here, we optimize the parameters $\theta_{(\cdot)}$ of the control policies $\pi_{\theta_1}$, $\pi_{\theta_2}$, and $\pi_{\theta_3}$ that map the vector of control parameters $\xi$, defined in \eqref{eq:dpc-xi}, to continuous-valued control variables $\dot m$, $T_\mathrm{e}$, and a slack variable $\tilde\delta\!\in\!\mathbb{R}^{M-1}$---representing the relaxed value of the binary control variable $\delta$. Notice that by fixing one of the binary variables, the dimension of $\tilde\delta$ is reduced to $M\!-\!1$. To compute control variables, we use separate neural modules that are represented by fully connected deep neural networks. Moreover, we sample the control parameters from a known probability distribution $P_\xi$ to emulate scenarios that are likely to occur during the system's operation, which defines a self-supervised nature of the DPC methodology. Functions $q(\cdot)$ and $p(\cdot)$ denote state and input penalty functions weighted by $\lambda_x$ and $\lambda_u$, respectively. Equation \eqref{eq:dpc-oneon} states that one chiller in the system must always remain in operation.

\subsection{Differentiable Rounding}
\label{sec:integrality}
To facilitate gradient-based optimization, each component of the problem \eqref{eq:MI-DPC} must be differentiable with well-defined gradients. This requirement is satisfied for all terms in \eqref{eq:MI-DPC} except for constraint \eqref{eq:dpc-binary}, which enforces the integrality of the binary variable $\delta$, thereby introducing non-differentiability. We enforce the binary integrality through a threshold-based discretization scheme with a fixed threshold of 0.5, given as
    \begin{align} \label{eq:unit-step}
\delta^{(i)}_{k} = & \begin{cases}
    1, & \text{if } \tilde\delta^{(i)}_{k}>0.5,\\[0pt]
0, & \text{otherwise.}\\[0pt]
\end{cases}
    \end{align}
This function behaves analogously to a Heaviside (unit step) function exhibiting a discontinuity at \num{0.5}. To enable gradient-based optimization, we leverage a Straight-through estimation heuristic introduced by \cite{bengio2013estimating}. We approximate the gradient of \eqref{eq:unit-step} using its differentiable surrogate, given by a scaled Sigmoid function
\begin{align}
\sigma(x)\!=\!\frac{1}{\left(1+e^{-\textcolor{black}{\mu}(x-0.5)} \right)}, 
\end{align}
where $\textcolor{black}{\mu}\!\in\!\mathbb{R}_{>0}$ is a slope parameter determining the sharpness of the transition, and the function is centered at \num{0.5}. Accordingly, the derivative is approximated as 
\begin{align}
    \frac{\partial \delta^{(i)}_{k}}{\partial \tilde{\delta}^{(i)}_{k}} 
    \approx \frac{\partial \sigma\!\left(\tilde{\delta}^{(i)}_{k}\right)}{\partial \tilde{\delta}^{(i)}_{k}}.
\end{align}

\subsection{Binary-Variance Regularization}
\begin{figure}[tb]
    \begin{center}
        \begingroup
       \footnotesize
       \resizebox{0.95\columnwidth}{!}{\input{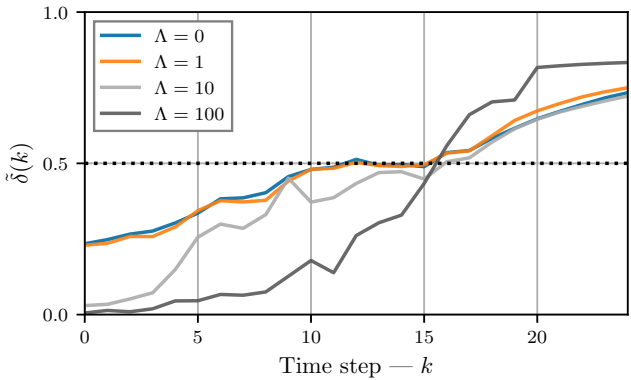}}
    \endgroup
    \end{center}
    \caption{\color{black}Effect of binary-variance regularization on relaxed integer values, showing switching behavior under a step change in cooling load from \num{150}\si{\kilo\watt} to \num{500}\si{\kilo\watt} occurring at time $k\!=\!20$ (setup from Sec.~\ref{sec:experiments}, $N\!=\!20$). The dotted black line indicates the rounding threshold.}
    \label{fig:BVR}
\end{figure}

To promote binary polarity of the relaxed variable $\tilde \delta$ and to complement the $\Delta\delta$ penalty in suppressing high-frequency switching, we introduce a binary-variance regularization (BVR) defined as  $ \|\tilde\delta^{(i)}(1-\tilde\delta^{(i)})\|_\Lambda^2,$
where $\Lambda\!\in\!\mathbb{R}^{M-1\times M-1}$ denotes the weighting matrix. As the weight increases, the regularization penalizes values of $\tilde\delta$ near the discretization threshold (\num{0.5}), as well as those outside the $[0,1]$ range. 
Based on extensive numerical experiments, this regularization has proven crucial to the proposed framework, as its omission leads to $\tilde\delta$ values clustering around the discretization threshold, thereby increasing the likelihood of high-frequency switching behavior. The influence of this regularization on the relaxed binary variable for different penalty magnitudes is illustrated in Figure~\ref{fig:BVR}.

\subsection{Mixed-Integer Policy Optimization}
The computational graph, illustrated in Figure~\ref{fig:MI-DPC}, comprises four components: a continuous neural policy, a differentiable rounding layer, a differentiable ordinary differential equation (ODE) solver (numerical integrator), and a model predictive loss function. Given synthetically sampled control parameters that include the initial conditions and an $N$-long vector of exogenous variables (server load), a closed-loop rollout of length $N$ is performed through the integrated system dynamics, constituting the single shooting forward pass. Consequently, the loss function is evaluated based on the resulting closed-loop trajectories. 

The backward pass is used to enable the policy optimization by propagating gradients through the computational graph. Specifically, gradients of the loss function with respect to policy parameters are obtained by back-propagating through the integrated system dynamics using either the backpropagation through time algorithm~\citep{puskorius1994} or the adjoint sensitivity method~\citep{Griewank}. The differentiability of the discretization layer and the numerical integrator ensures that the gradient signals are transmitted continuously through closed-loop trajectories. These gradients are then used to adjust the policy parameters in the direction of decreasing loss.
Consequently, the policy parameters are iteratively updated.
The high-dimensional and non-convex nature of the underlying parameter space necessitates the use of stochastic gradient descent (SGD), for which global optimality cannot be guaranteed.

\section{Numerical Experiments}
\label{sec:experiments}
In this section, we present and discuss the results of a numerical study in which we test the three control strategies. The control strategies are tested for systems comprising different numbers of chillers and various prediction horizon lengths. Each system consists of identical chiller units with parameters listed in Tab.~\ref{tab:chiller-parameters}. 
\begin{table}[b] 
\centering
\caption{Chiller system parameters.}
\label{tab:chiller-parameters}
\setlength{\tabcolsep}{3.5pt} 
\renewcommand{\arraystretch}{1.0} 
\begin{tabular}{llcc} 
\toprule
\text{Symbol} & \text{Description} & \text{Value}   & \text{Unit}\\
\midrule
$c_p$ & Specific heat of water & \num{4.184}  & \si{\kilo\joule\per{\kilo\gram\celsius}} \\
$C$ & Thermal capacitance & \num{14644}  & \si{\kilo\joule\per{\celsius}} \\
$C_r$ & Thermal capacitance & \num{29288}  & \si{\kilo\joule\per{\celsius}} \\
$\rho$ & Base chiller power & \num{10}  & \si{\kilo\watt} \\
$Q_{\max}$ & Maximum rated cooling & \num{500}  & \si{\kilo\watt} \\
$a_0$ & Efficiency coefficient & \num{1}  & \si{-} \\
$a_1$ & Efficiency coefficient & \num{19.33}  & \si{-} \\
$a_2$ & Efficiency coefficient & \num{-18.33}  & \si{-} \\
$\eta_\mathrm{s}$ & Heat exchanger efficiency & \num{0.7}  & \si{-} \\
$\eta_\mathrm{r}$ & Heat exchanger efficiency & \num{0.75}  & \si{-} \\
$\gamma$ & Pump power coefficient & 9.62e-4  & \si{\kilo\watt\second\cubed\per{\kilo\gram\cubed}} \\
$\dot{m}^\text{min}$ & Minimum mass flow rate & \num{5}  & \si{\kilo\gram\per{\second}}\\
$\dot{m}^\text{max}$ & Maximum mass flow rate & \num{20}  & \si{\kilo\gram\per{\second}} \\
$T_\text{s}^{\text{min}}, T_\text{e}^{\text{min}}, T_\text{r}^{\text{min}}$ & Minimal temperatures & \num{8}  & \si{\celsius}\\
$T_\text{s}^{\text{max}}, T_\text{e}^{\text{max}}$ & Maximal temperatures & \num{12}  & \si{\celsius}\\
$T_\text{r}^{\text{max}}$ & Maximal return temp. & \num{40}  & \si{\celsius}\\
\bottomrule
\end{tabular}
\end{table}

\subsection{Experimental Setup}

Throughout the study, the system dynamics were simulated using a constant time step length of $\Delta t\!=\!180$ \si{\second} and a finite impulse response filter with coefficients $h_l\!=\![0.45,0.2,0.15,0.1,0.05,0.05]$. The load signal for each experiment was algorithmically generated to emulate daily variations in server load. Specifically, the nighttime load amplitude was sampled from a uniform distribution $\mathcal{U}\left(100\si{\kilo\watt},350\si{\kilo\watt}\right)$, while the daytime amplitude was sampled from $\mathcal{U}(300\si{\kilo\watt},0.75 \sum_iQ^{(i)}_\text{max})$ with additive white noise and a transition period of \num{4} \si{\hour}. The system states were uniformly initialized within their respective operating ranges $T_\mathrm{r}\!\sim\!\mathcal{U}\left(T^{\min}_\mathrm{r}, T^{\max}_\mathrm{r}\right)$ and $T_\mathrm{s}\!\sim\!\mathcal{U}\left(T^{\min}_\mathrm{s}, T^\mathrm{max}_\mathrm{s}\right)$. Each neural control policy module employed an identical architecture, consisting of three hidden layers of size \num{200} with ReLU activations and $[0,1]$ normalization on the input layer. The control loss weights were manually finetuned to $\Lambda\!=\!200$, $\lambda_x,\lambda_u\!=\!10$, $\textcolor{black}{\lambda_W}\!=\!0.001$, $R\!=\!20$. \textcolor{black}{It is important to emphasize that the magnitude of $\Lambda$ critically influences the behavior of the proposed framework. Specifically, lower values of $\Lambda$ led to instances of high-frequency switching, whereas excessively large values resulted in training instability.}
Furthermore, the hyperparameters were also manually adjusted a learning rate of \num{0.006}, a batch size of \num{10000}, with training and development dataset sizes of \num{30000} and \num{10000}, respectively. \textcolor{black}{The learning rate was selected to ensure stable optimization, while the relatively large batch size was chosen to provide reliable gradient estimates.}
\textcolor{black}{To further mitigate the risk of gradient explosion during optimization, the gradient norms were clipped with a maximum allowed norm of \num{100}, such that all gradients $\{g_i\}$ were uniformly rescaled to satisfy $\|\mathbf{g}\|_2\le100$.} The sigmoid slope coefficient was kept at the default value $\textcolor{black}{\mu}\!=\!1$, as with scaling values, an excessively conservative binary behavior was exhibited, thereby limiting actuation flexibility and reducing potential energy savings. In inference the continuous control inputs of MI-DPC policies were projected onto their feasible sets via elementwise clipping.

The numerical experiments were conducted using the NeuroMANCER v1.5.6 library~\citep{Neuromancer2023} running with PyTorch v2.8.0 and CUDA 12.8 with MI-DPC. The MI-MPC formulation was implemented in Pyomo v6.9.4~\citep{bynum2021pyomo} and solved using Gurobi v12.0.3~\citep{gurobi}. 
\textcolor{black}{Similarly to \citep{pan2024nonlinear}, the nonconvexity arising from bilinear coupling between continuous and binary decision variables within the heat-transfer equations, the problem was reformulated using McCormick envelope relaxations---resulting in a mixed-integer quadratic programming approximation solvable by Gurobi.} In addition, higher-order nonlinearities, such as cubic term in \eqref{eq:pump_power}, were similarly reformulated using slack variables and McCormick relaxation. Nevertheless, the MI-MPC formulation remains computationally intractable for real-time applications, as discussed in the following Section.

\begin{figure*}[htb!]
    \begin{center}
        \begingroup
       \footnotesize
       \resizebox{\textwidth}{!}{\input{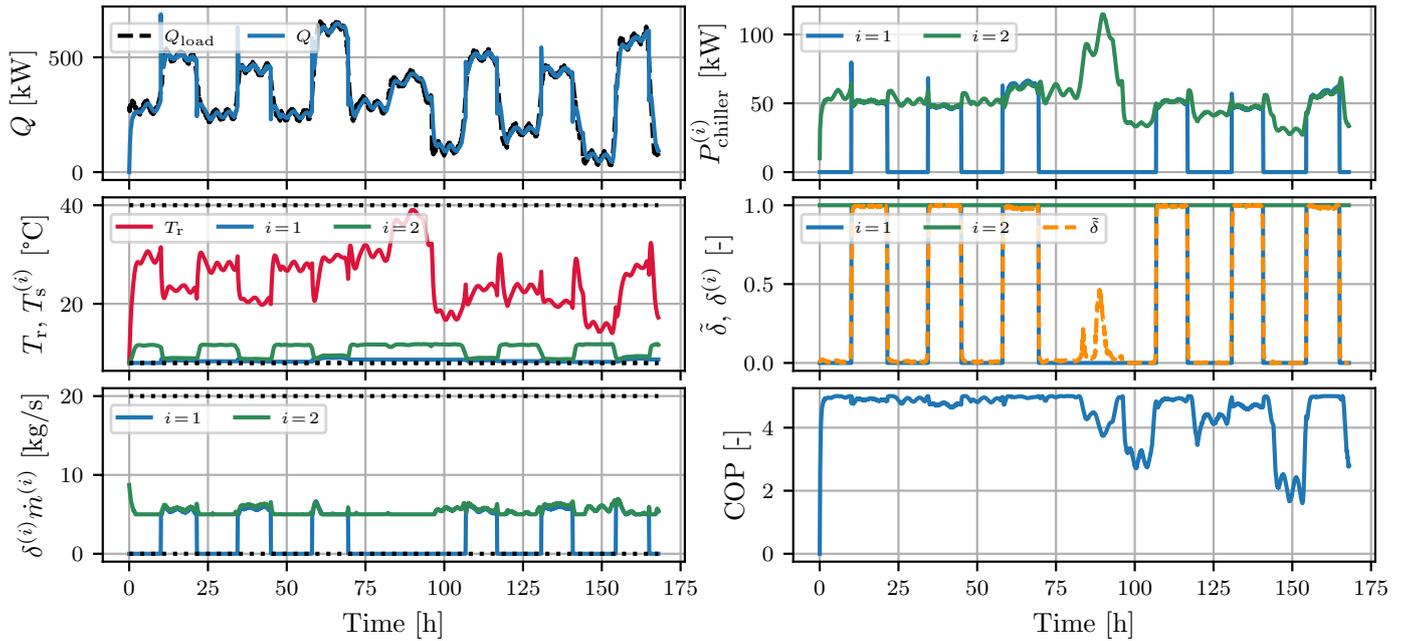}}
    \endgroup
    \end{center}
    \caption{Closed-loop seven-day simulation results of a two-chiller system obtained with MI-DPC for a prediction horizon of $N\!=15$.}
    \label{fig:control}
\end{figure*}

\textcolor{black}{
The closed-loop simulation of MI-DPC, implemented in a receding-horizon manner, is shown in Figure~\ref{fig:control}. The observed increase in chiller power consumption between $80$\si{\hour} and $90$\si{\hour} is attributed to the PLR-COP characteristic. During this interval, the chiller's operating point shifts toward a less efficient PLR range, resulting in a lower COP and, consequently, increased power consumption despite comparable thermal load values. In addition, the controller shows a tendency to activate an additional chiller within this time interval, as reflected by the rising value of the relaxed integer variable $\tilde \delta$. However, $\tilde \delta$ remains below the rounding threshold, as activating an additional unit would likely increase the load-tracking error, thereby outweighing the potential improvement of energy efficiency within the objective function.  
}

\subsection{Computational Scalability}
\begin{figure}[t]
    \begin{center}
        \begingroup
       \footnotesize
       \resizebox{\columnwidth}{!}{\input{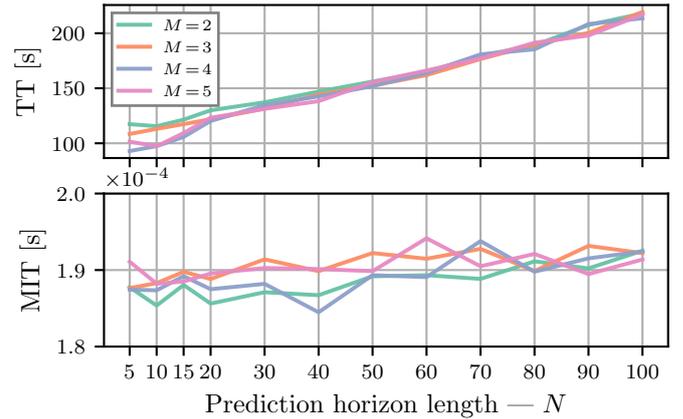}}
    \endgroup
    \end{center}
    \caption{\color{black}Computational scalability of MI-DPC across different number of chillers ($M$) and horizon lengths ($N$). The top panel illustrates the total training time (TT), while the bottom panel reports the mean inference time (MIT).}
    \label{fig:times}
\end{figure}

Table~\ref{tab:energy_cop} summarizes the simulation results obtained with the RBC and MI-DPC policies obtained from a simulation length of \num{7} days. Note that we do not include MI-MPC results due to computational intractability; instead, we report the mean inference time, with the maximum solver time limited to \num{180} \si{\second}---a practically tractable upper bound for real-time implementation, equivalent to the control sampling period. This highlights the notorious scalability limitations of hard mixed-integer problems. In contrast, the MI-DPC approach alleviates computational scalability issues in both offline and online stages. As shown in Fig.~\ref{fig:times}, the mean inference time remains nearly constant even as the prediction horizon increases, since the same network architecture is used across all scenarios. The number of trainable parameters increases only with the input dimension, which grows with the prediction horizon to accommodate the longer preview of the load variable. When it comes to the training time (TT), scalability with respect to the prediction horizon length exhibits an approximately linear trend. 
\begin{table}[b]
\color{black}
\caption{Summary of seven-day simulation results for the RBC, MI-DPC, and MI-MPC implementations for different numbers of chillers ($M$) and prediction horizon lengths ($N$).}
\label{tab:energy_cop}
\setlength{\tabcolsep}{0.75pt}
\scriptsize
\begin{tabular}{@{}llcccccc@{}}

    \toprule
    \multirow{2}{*}{Method} & \multirow{2}{*}{Metric} & \multicolumn{3}{c}{$M\!=\!2$} & \multicolumn{3}{c}{$M\!=\!3$} \\
    \cmidrule(lr){3-5} \cmidrule(lr){6-8}
    &  & $N\!=\!5$ & $N\!=\!10$ & $N\!=\!15$ & $N\!=\!5$ & $N\!=\!10$ & $N\!=\!15$ \\

\midrule
\multirow{6}{*}{RBC} & EC [MWh] & 13.52 & - & - & 17.53 & - & - \\
 & EC Chillers [MWh] & 13.23 & - & - & 17.14 & - & - \\
 & EC Pumps [MWh] & 0.28 & - & - & 0.39 & - & - \\
 & COP [-] & 4.08 & - & - & 3.88 & - & - \\
 & Num. of switches [-] & 9 & - & - & 22 & - & - \\
 & Mean RCE [\%] & 5.38 & - & - & 5.90 & - & - \\
\midrule\multirow{10}{*}{MIDPC} & EC [MWh] & 12.39 & 12.16 & 12.24 & 15.96 & 15.64 & 15.57 \\
 & Savings [\%] & 8.38 & 10.01 & 9.45 & 8.95 & 10.79 & 11.17 \\
 & EC Chillers [MWh] & 12.34 & 12.13 & 12.20 & 15.87 & 15.55 & 15.52 \\
 & EC Pumps [MWh] & 0.04 & 0.04 & 0.04 & 0.09 & 0.08 & 0.05 \\
 & COP [-] & 4.44 & 4.54 & 4.51 & 4.26 & 4.37 & 4.38 \\
 & Num. of switches [-] & 15 & 14 & 12 & 30 & 30 & 28 \\
 & Mean RCE [\%] & 7.21 & 8.49 & 8.83 & 7.39 & 7.72 & 9.31 \\
 & MIT [s] & 1.9e-04 & 1.9e-04 & 1.9e-04 & 1.9e-04 & 1.9e-04 & 1.9e-04 \\
 & TT [s] & 117.36 & 115.53 & 121.47 & 108.46 & 113.10 & 117.40 \\
 & NTP [-] & 248205 & 251205 & 254205 & 249408 & 252408 & 255408 \\
\midrule MIMPC & Inference Time & 157.44 & 181.33 & 181.70 & 182.77 & 181.74 & 180.98 \\
\bottomrule
\end{tabular}
\begin{tablenotes}
\scriptsize
\item Used acronyms: EC---Energy Consumption, EC-COP---Effective Chiller Coefficient of Performance, MIT---Mean Inference Time, NTP---Number of Trainable Parameters, TT---Training Time, RCE---Relative Control Error.
\item[1] Measured using a single thread of Intel i9-14900KF.
\item[2] Trained using an NVIDIA GeForce RTX 5090.
\item[3] Solved with the gurobi solver.
\end{tablenotes}
\end{table}

As evidenced by the simulation results, the principal advantage of MI-DPC lies in its computational efficiency during both the controller synthesis phase and inference. The method achieves inference times several orders of magnitude faster than the implicit MI-MPC formulation, thereby potentially enabling solving the mixed-integer nonlinear optimal control problems on computationally constrained edge devices in real time---a capability not attainable with conventional approaches.

\subsection{Performance Comparison}
In comparison to the RBC, the proposed MI-DPC framework achieved up to \num{11.17} \% energy savings. These savings resulted primarily from more adaptive and frequent chiller staging---albeit at the expense of a bigger relative load tracking error, as well as from reduced water pump power consumption. With MI-DPC we are able to simultaneously schedule chiller activation and adjust the continuous control variables in an adaptive manner.
The RBC was implemented with staging thresholds of $\underline{t}\!=\!0.15$, $\overline{t}\!=\!0.6$ and constant mass flows and evaporation temperatures of $\dot{m}\!=\!10$ \si{\kilo\gram\per\second} and $T_\mathrm{e}\!=\!10$ \si{\celsius}, respectively.

In Table~\ref{tab:energy_cop}, the Effective Chiller COP (EC-COP) is computed as 
\[
\text{EC-COP}=\frac{\sum_{k=0}^{N_\text{s}}\sum_{i=1}^MQ^{(i)}_k}{\sum_{k=0}^{N_\text{s}}\sum_{i=1}^MP_{\text{chiller,k}}^{(i)}},
\]
where $N_\text{s}$ is the length of simulation. The Relative Control Error (RCE) is computed as 
\[
\text{RCE}_{k}=\frac{|Q_{\text{load,}k} - \sum_{i=1}^MQ^{i}_{k}|}{Q_{\text{load},k}} \cdot100\%.
\]

\begin{figure}[htb]
    \begin{center}
        \begingroup
       \footnotesize
       \resizebox{\columnwidth}{!}{\input{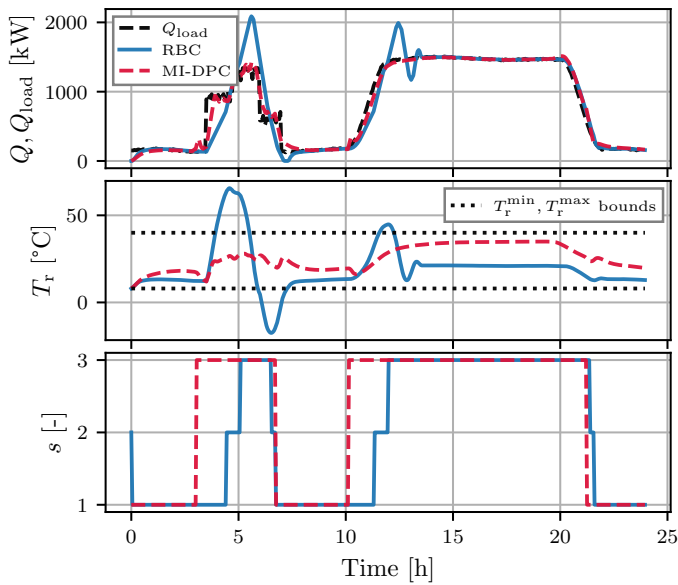}}
    \endgroup
    \end{center}
    \caption{\color{black}Closed-loop simulation results of a chiller plant ($M\!=\!3$, $Q_{\max}\!=\!1$ \si{\mega\watt}) with MI-DPC and RBC policies, highlighting the importance of predictive action for stable operation, when considering cooling ramp-rate constraints.}
    \label{fig:violation}
\end{figure}

The significance of the predictive capability of MI-DPC became evident through extensive numerical experiments. In addition to energy savings, the MI-DPC framework outperformed the RBC in terms of process constraints. Specifically, when considering cooling ramp-rate constraints, i.e., $\dot{Q}_{\min}\leq\!\dot{Q}\!\leq\dot{Q}_{\max}$, which have been observed in real chiller plant operations and reported in~\citep{shan2021controlling}; the RBC may fail to activate chillers in time when encountering a steep increase in cooling demand. As illustrated in Figure~\ref{fig:violation}, the RBC stages chillers solely based on the staging threshold. However, due to the cooling ramp limitation, the return temperature can exceed the process safety bounds. \textcolor{black}{ In contrast, the MI-DPC framework anticipates future load changes and stages chillers proactively based on the known future cooling demand. In the experiment, during the time interval from $3$\si{\hour} to $7$\si{\hour}, the results demonstrate that MI-DPC effectively adapts to abrupt changes in the cooling load profile, even when such changes were not represented in the training dataset.}

\textcolor{black}{
Moreover, extensive simulation experiments indicate that extending the prediction horizon beyond $N\!=\!15$ does not provide additional energy savings for the considered system. This behavior can be attributed primarily to the relatively fast thermal dynamics of the modeled chiller plant, which allows for effective control within relatively shorter horizon lengths. Extending the prediction horizon length may be advantageous for systems with slower dynamics, such as those including thermal storage tanks or larger thermal masses. From a computational standpoint, the MI-DPC framework maintains scalability even as the prediction horizon length increases, as illustrated in Figure~\ref{fig:times}.
}

\textcolor{black}{
\subsection{Practical Challenges and Limitations}
In this study, we assume a fully white-box, deterministic system model for policy synthesis and testing, providing a controlled setting to evaluate and benchmark the MI-DPC framework. In practical applications, model parameters must be identified from real plant data and model fidelity validated to ensure robustness and generalizability. This can be addressed through grey-box parameter estimation, where the system parameters are estimated using, for instance, a nonlinear least-squares, or physics-informed modeling strategies~\citep{UDEs2021,koch2025}. However, practical MPC deployments are further challenged by model–plant mismatch, unmodeled dynamics, sensor and actuator noise, and actuator delays, which may degrade closed-loop performance and compromise constraint satisfaction if not explicitly accounted for. Current MPC frameworks address these issues through disturbance modeling, state estimation, uncertainty propagation, and constraint tightening strategies~\citep{qin2003survey,bemporad2007robust}.
}%

\textcolor{black}{
It is further recognized that the problem is solved approximately, without a guaranteed global optimum. Moreover, while integrality and input constraints satisfaction are enforced through projection-based methods, strict feasibility for state constraints is not guaranteed. These aspects will be addressed in future work.
}

\section{Conclusions}

We presented a mixed-integer nonlinear Differentiable Predictive Control (MI-DPC) framework for real-time optimization of multi-chiller plants, unifying discrete staging and continuous setpoint control within a single differentiable program. The approach leverages (i) differentiable rounding with straight-through estimator gradients to handle binary decisions, (ii) binary-variance regularization to suppress chattering, (iii) a nonlinear chiller plant dynamics solver with a differentiable ODE solver, and (iv) a model-predictive loss evaluated over rollout trajectories, yielding an explicit neural policy with constant-time inference. In simulations on multi-chiller systems, MI-DPC achieved up to $11\%$  energy savings relative to an industrial RBC baseline while preserving process constraints, and delivered orders-of-magnitude faster inference than MI-MPC, indicating strong potential for a practical and scalable deployment on large-scale HVAC operations.

Building on the findings of this study, several extensions are envisioned.
The chiller plant model is to be extended to include cooling towers, allowing for further energy savings potential through optimization of condenser-water loop setpoints~\citep{huang2017improved, wang2019cooling}. The integration of water-side economizers can also be explored, as these have been shown to reduce energy through free-cooling~\citep{faulkner2025development} and introduce additional discrete control variables for economizer activation.
Future work will also incorporate probabilistic cooling-load forecasting to more accurately capture real-world uncertainty.
Finally, the framework may be adopted for experimental validation in real-world chiller plants.




\section*{DECLARATION OF GENERATIVE AI AND AI-ASSISTED TECHNOLOGIES IN THE WRITING PROCESS}
No generative AI and AI-assisted technologies were used in the writing process.

\bibliography{ifacconf}             
                                                   







\appendix
\end{document}